\documentclass{aa}
\usepackage{graphicx}
\begin{document}
\title{RR Lyrae stars in the inner LMC: Where did they form?}
\author{Annapurni Subramaniam\inst{1} \& Smitha Subramanian\inst{1,2}}
\offprints{Annapurni Subramaniam}
\institute{Indian Institute of Astrophysics, Koramangala II Block, Bangalore - 34\\
           Department of Physics, Calicut University, Calicut, Kerala\\
            \email{purni@iiap.res.in, smitha@iiap.res.in}}
\date{Received / accepted}
\titlerunning{RR Lyrae stars in the LMC}
\authorrunning{Subramaniam \& Subramanian}
\abstract
{RR Lyrae stars (RRLS) belong to population II and are generally used as a tracer of the
host galaxy halo.}
{
The surface as well as vertical distribution of RRLS in the inner Large Magellanic Cloud (LMC) are studied to understand 
whether these stars are actually formed in the halo.
}
{RRLS identified by the OGLE III survey 
are used to estimate their number density distribution.
The scale-height of their distribution
is estimated using extinction corrected average magnitudes of ab type stars. }
{The density distribution mimics the bar, confirming results in the literature.
The distribution of their scale height
indicates that there may be two populations, one with smaller scale-height, very similar
to the red clump stars and the other, much larger. The distribution of the reddening-corrected
magnitude along the minor axis shows variation, suggesting an inclination. The inclination
is estimated to be $ i = 31.3 \pm 3^o.5 $ degrees, very similar to the inclination of the disk.
Thus, the RRLS in the inner LMC mimic the bar and inclination of the disk, suggesting that a major
fraction of RRLS is formed in the disk of the LMC.
}
{
The results indicate that the RRLS in the inner LMC
trace the disk and probably the inner halo. They do not trace the extended metal-poor halo of the LMC.
We suggest that a major star formation event happened in the LMC
at 10-12 Gyrs ago, resulting in the formation of most of the inner
RRLS, as well as probably the globular clusters, inner halo and the disk of the LMC. 
}
\maketitle
\section{Introduction}
The Large  Magellanic Cloud (LMC) is known to be a disk galaxy with or without a halo.
There have been many efforts to find evidence for the presence of a halo in the LMC.
Such evidence is looked for in the old stellar population, such as globular clusters (GCs) and
field RR Lyrae stars (RRLS). The oldest GCs appear to lie in a flat rotating disk whose
velocity dispersion is 24 km s$^{-1}$ (\cite{k91} and \cite{sch92}).
\cite{vb04} re-estimated the velocity dispersion of the GCs and concluded that they
could still have formed in the halo. Since the number of GCs in the LMC is small 
(13 clusters), it is difficult to infer the signature of the halo from this
sample. Another tracer is the RRLS, which are almost as old as the oldest GCs.
Recent surveys of the LMC such as OGLE and MACHO have identified a
large number of RRLS. Among the follow up studies, \cite{m03} found kinematic 
evidence for the
LMC halo, by estimating the velocity dispersion in the population of RRLS. \cite{al04} estimated
the mass of the LMC based on RRLS surface density and remarked that their
exponential scale length is very similar to that of the young LMC disk. 
\cite{f99} remarked that the similarity of their scale length
to that of the young stellar disk suggests that 
the RRLS in the LMC are disk objects, like old clusters, 
and supported the view that the LMC may not have a metal-poor halo.

\cite{as06} studied the density distribution and found that it is elongated as the LMC bar. Since these
stars were found to have a disk-like distribution and halo-like location, she speculated that
these stars might have formed in the disk and its present location could be due to later mergers.
The OGLE III catalogue of RRLS, (\cite{sz09}) confirmed this bar-like elongation.
In this study, we used the RRLS catalogue of \cite{sz09} in the LMC, consisting of
17693 ab type RRLS to obtain their number density and scale height distributions. The method used by
\cite{as06} is followed here. 
The density distribution is clearly found to be elongated
and the PA estimate matches that of the RC stars. 
The scale height found by \cite{as06} was larger than the scale height
for the disk as delineated by red clump (RC) stars. Thus it was suggested that the inflated distribution
of the RRLS could be due to mergers and the RRLS might have initially formed in the disk. 
The scale-height distribution of the OGLE III RRLS is estimated and is compared with that obtained from the
RC stars by \cite{ss09a}. The comparison identifies two populations, where one 
population is found to 
follow the disk population. We have explored whether another property of the LMC disk, i.e., inclination, is
present in the RRLS. The structure of the paper is as follows: the number density distribution is estimated in 
section 2, the scale-height distribution is estimated and compared with the RC stars in section 3, the
inclination of the RRLS is estimated in section 4 and is followed by a discussion in section 5.

\begin{figure}
\resizebox{\hsize}{!}{\includegraphics{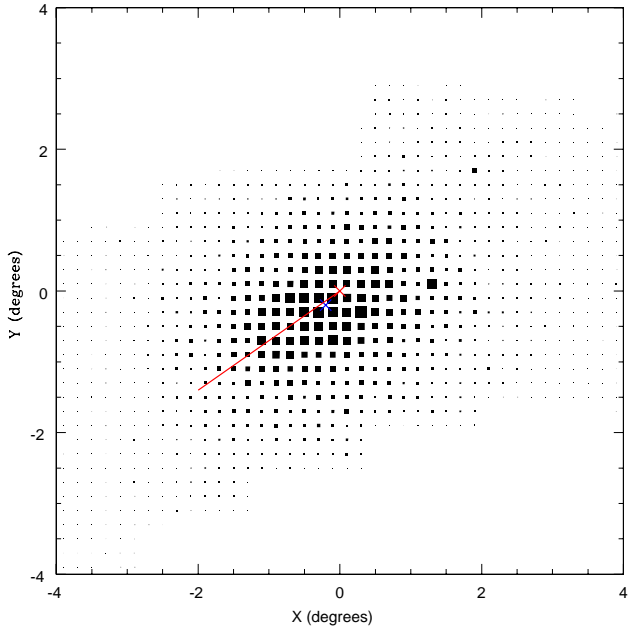}}
\caption{ Number density distribution of RR Lyrae stars in the LMC. The optical center is shown
as red cross, the center of the RRLS density is shown in blue cross, the PA$_{maj}$ is shown
in red line.
\label{fig1}}
\end{figure}

\section{Estimation of the number density distribution of RRLS}
\cite{sz09} presented a catalogue of RRLS discovered in the
inner parts of the LMC, consisting of 17693 objects. 
The center of the LMC was taken as $\alpha$ = 05$^h$19$^m$38$^s$; 
$\delta$ = $-$69$^o$27'5".2 (\cite{df73}). The $\alpha$ and $\delta$ were converted to 
the projected X and Y
coordinates. The data are binned in 0.2 degree bins on both the axes and 
the number of RRLS in each box is counted to obtain the number
density in square degrees. A plot of their density is shown in 
figure~\ref{fig1}, where the variation in the number density is clearly seen.
On the whole, the smooth spatial variation in the density
is found to have an elongated distribution. The direction of elongation is found to
be similar to the elongation of the bar of the LMC. 
The position angle (PA) of the elongation 
is estimated to be PA = 125 $\pm$ 17$^o$ for the eastern side and this value is within the errors of the value
found by \cite{as06} using OGLE II RRLS ( PA = 112.5$^o$ $\pm$ 15.$^o$3) 
and by \cite{ps09} ( PA = 112.4$^o$). 
 A clear value for the
PA emerges on the eastern side, but a clear value does not emerge on the western side.
The above value is very similar to the PA$_{maj}$ as estimated from RC stars, 114$^o$.4 $\pm$ 22$^o$.5, (\cite{S04}) and also the PA$_{maj}$ from the red giants 122$^o$.5 $\pm$ 8$^o$.3 (\cite{v01}). 
All the above values are the same within the errors.
\cite{as06} also found that the density distribution of RRLS is similar to the RC and red giant stars
along the major axis.
The density distribution of the RRLS is found to be located slightly away
from the adopted optical center of the LMC. Their center was 
found to be
$\alpha$ = 05$^h$ 20$^m$.4$\pm$0.$^m$4 and $\delta$ = $-$69$^o$ 48'$\pm$5' by \cite{as06}. We also
find the center of the RRLS to be same as above, within errors. This is also similar to the value
estimated by \cite{al04}. This center is denoted in figure 1
as blue cross, and is slightly shifted with respect to the optical center.

Thus, the bar-like elongation seen in the RRLS is significant. It is not possible for the RRLS to have formed
in the halo and still have a bar like elongation. 
A probable scenario is that most of the RRLS are formed in the disk.
To summarise, the RRLS in the inner LMC
are likely to have formed in the disk and are unlikely to trace the halo.

\begin{figure}
\resizebox{\hsize}{!}{\includegraphics{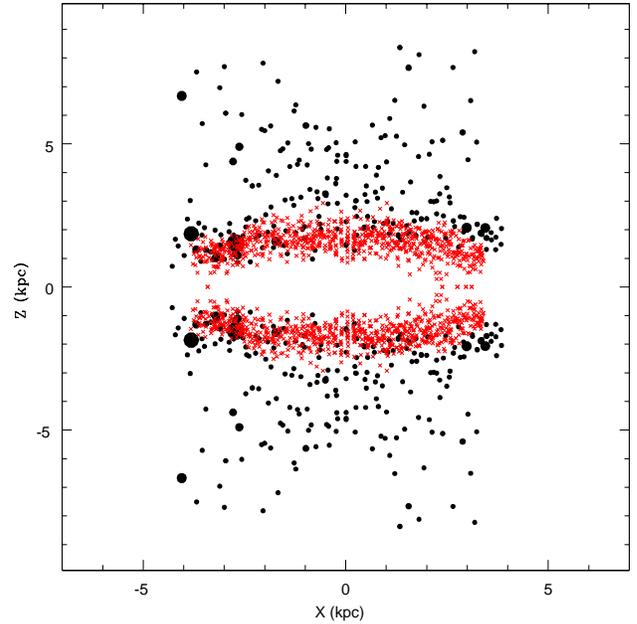}}
\caption{ The distribution of scale-height of ab type RRLS (black) 
compared with that of RC stars (red). 
The scale heights shown are the dispersion in the mean magnitude reflected about z=0.
The larger sized dots indicate larger error in the estimated dispersion. 
\label{fig2}}
\end{figure}
\section{Estimation of dispersion in the mean magnitude of ab type RRLS}
The ab type stars could be considered
to belong to a similar sub class and hence assumed to have similar properties.
The mean magnitude of these stars in the I pass band, after correcting for the metallicity and extinction
effects, can be used for the estimation of distance. On the other hand, the observed
dispersion in their mean magnitude is a measure of the depth in their distribution.
\cite{ss09b} presented a high-resolution reddening map of the region
observed by the OGLE III survey. This data is used to estimate the extinction to individual
RRLS. Stars within each bin of the reddening map are assigned a single reddening
and it is assumed that the reddening does not vary much within the bin (4.44 X 4.44 sq. arc min). The contribution
to the dispersion from the variable extinction is minimised, though there may still be a
non-zero contribution to the estimated dispersion.


The observed distribution of ab type stars
in the X and Y direction are binned such that there are more than 10 stars in most of the locations and
a maximum of $\sim$ 300 stars in some locations. We used a bin size of 0.$^o$5 in X
and 0.$^o$5 in Y. We estimated the mean magnitude and dispersion for 231 locations.
This value of the dispersion can be used to estimate the scale-height, but
has contributions from (1) photometric errors, (2) range in
the metallicity of stars, (3) intrinsic variation in the luminosity due to evolutionary
effects within the sample and (4) the actual depth
in the distribution of the stars (\cite{C03}). We need to remove the
contribution of the first three terms ($\sigma_{(int)}$) so that the value of the
last term ($\sigma_{(dep)}$) can be evaluated.
\cite{C03} estimated the value of $\sigma_{(int)}$ as 0.1 mag for their sample. 
\cite{as06} estimated a value of 0.15 mag for a similar analysis based on OGLE II data of RRLS.
When we tried to subtract a value of 0.15 mag, we found that many regions have a dispersion less than this
value, especially at locations away from the center. Hence we decided to ignore this reduction 
and use the total value of dispersion to estimate the scale height, very well aware that values estimated
will be upper limits. Thus we do not derive any quantitative estimates on scale height.
We have plotted the observed dispersion in magnitude (bins with less than 10 stars are excluded),
converted into Z distance in an edge-on view along the X-axis shown in Figure 2. The scale
height displayed is the Gaussian dispersion of z-heights in each
bin, rather than the conventional scale height of an exponential
distribution.
 The distance is obtained
from the relation 0.1 mag = 2.3 kpc. In order to give a complete edge-on view, the estimated
depth in magnitude is halved and shown along the +ve and -ve Z-axis. In order to compare this depth with
the disk population, we have shown the RC depth (halved for the edge-on view) estimated by \cite{ss09a}. Their result also points to a thicker LMC disk, even after correcting for inclination, as shown by
the red points.
The figure clearly shows two population of RRLS, one population being very much like the RC stars in depth
distribution and the other showing an inflated distribution. The number of regions identified with the disk-like
RC stars increases away from the center. The inflated regions are located more or less in the bar region. The
figure also suggests an X-shaped distribution in the edge-on view. An X-Y view of the dispersion
is also shown as an inset in figure 3, where the size of the points is proportional to the dispersion. 
This might give clues to the formation of this
distribution, if the RRLS are actually formed in the bar/disk and then became inflated later. 

\begin{figure}
\resizebox{\hsize}{!}{\includegraphics{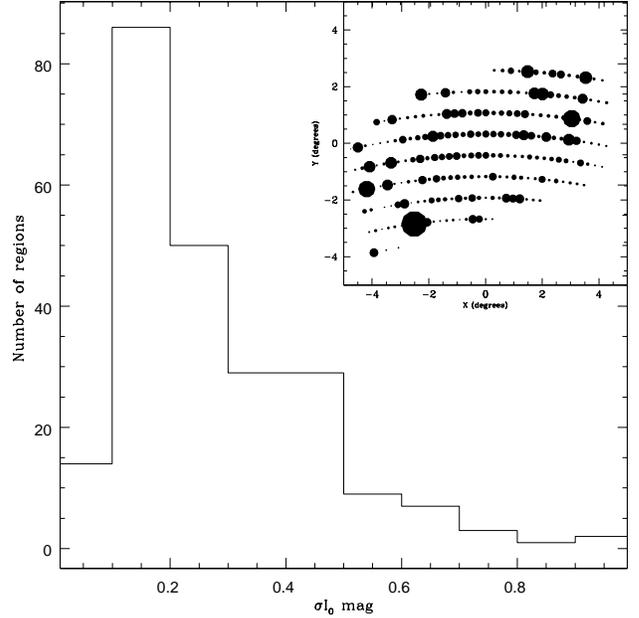}}
\caption{ The histogram of the dispersion in RRLS ($\sigma I_0$). 
The figure in the inset shows the X-Y location of the
bins, where the size of the points is proportional to the dispersion.
\label{fig3}}
\end{figure}
In order to estimate the fraction of locations of each population, the histogram of $\sigma I_0$ is shown in Figure 3.
The histogram has a peak in the 0.1 - 0.2 mag bin, where 43\% (100 regions out of 231) of the locations 
show a dispersion 
of less than 0.2 mag. This explains why we could not subtract the estimated contribution of $\sigma_{(int)}$. These regions are located all over, with increasing number away from the bar. 47\% of regions show
dispersions between 0.2 - 0.5 mag, and most of them are located within the bar region. About 10\% of regions show
dispersions of more than 0.5 mag. This clearly shows out that majority of the RRLS are in two types of distribution,
one disk-like and the other with a larger scale-height. This analysis might suggest that not more than 10\% of the 
RRLS in the inner LMC are likely to have formed in the extended halo.
                                                                                                                    
\begin{figure}
\resizebox{\hsize}{!}{\includegraphics{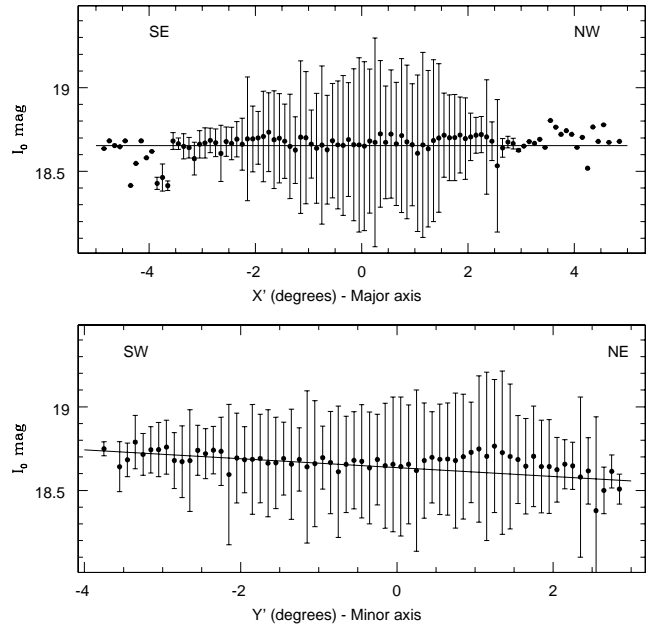}}
\caption{ The edge on view of the binned $I_0$ magnitudes along the major and minor axes. The direction of
the inclination is shown in the lower figure.
\label{fig4}}
\end{figure}
\section {Inclination of the distribution of RRLS}
Having estimated the major axis of the elongation, we shifted the coordinate system from X-Y to X'-Y' such that
the new X' and Y' are along the major and minor axes respectively. In order to see the disk+bar of the RRLS in the
edge on view, along the major and minor axes, we have shown the binned $I_0$ mag (binned between 18.0 - 19.0) 
against the two axes in figure 4.
The almost horizontal distribution along the major axis confirms that the estimated PA is close to the line
of nodes.
The bar region was excluded (X' = Y' = 2$^o$ degrees), and the remaining 57 data points were fitted to
estimate the slope as 0.0099$\pm$0.0028 mag/degree, the mean $I_0$ mag (y-intercept) as 18.654, with a correlation
coefficient of 0.43. The low correlation coefficient is due to the extra planar features seen beyond 3 degrees from the 
center. The estimated fit is shown in the figure, it fits the inner regions as well. 
 
The edge-on view along the minor axis shows a variation in $I_0$ magnitude, which suggests inclination
of the plane containing the RRLS. We excluded the regions between 0 and 2.$^o$0 of the minor axis, where significant
deviation is noticed. We fitted the rest of the regions
and the inclination was estimated. The values estimated using 46 points are 
inclination (slope) = -0.0265$\pm$0.0036 mag/degree, mean magnitude (y-intercept) = 18.637 mag and the correlation
coefficient was found to be 0.74. The fit as shown in the figure reveals that the fit is satisfactory, except for the
extra-planar feature towards the north eastern part. The inclination estimated here, $i = 31.3\pm3.^o5$, 
is very similar to the previous estimates of inclination for the disk 
( $i= 30.7\pm1.^o1$ \cite{n04}, $i=34.7\pm6.^o2$ 
\cite{vc01} ). If we consider all the points, including the extra-planer features, the
inclination becomes $i = 20.8\pm3.^o5$, using 66 points. The extra-planer feature, suggestive of an
extension behind the disk, is located close to
the 30 Doradus star forming region. In this analysis, we have considered reddening estimated using RC stars. 
This feature could arise because of uncorrected reddening for at least part of the RRLS in this region.
The colour of the dereddened RRLS were found to be relatively redder in these locations, supporting the
above idea.
To summarise, we establish the significant result that the RRLS
distribution has a line of nodes as well as an inclination similar to that of the LMC disk, 
which confirms the idea that a major
fraction of RRLS is in a non-spheroidal distribution.

\section{ Discussion: Clues to the early formation of the LMC disk}
The distribution of RRLS in the inner LMC is found have a major axis, line of nodes and inclination
similar to the disk. These three measures suggest that the RRLS are not in a spheroidal system, but
on an equatorial plane, similar to that of the LMC disk.
Also, this is the oldest tracer that shows disk properties. Therefore, turning the
argument around, we suggest that RRLS trace the early formation of the disk in the LMC.
The bar feature in the RRLS could be either due to them forming in the bar or being trapped by the bar, 
after they formed. 
The oldest disk as traced by the RRLS and the youngest disk
traced by the Cepheid (\cite{n04}) have similar inclinations. With respect to the
scale-height, we estimate that 
less than half of the stars follow the disk scale-height, whereas a similar fraction follow an inflated  
distribution. Only a small fraction ($\le$ 10\%) is likely to trace the extended halo
of the LMC. We suggest the following model for the formation of RRLS and the early evolution of the LMC.

The LMC showed a prominent star formation event at the time of the formation of RRLS, which mimics the LMC disk.
The enhanced density of RRLS near the central region is suggestive of a prominent star formation event.
This probably is the oldest and most significant star forming event, with the result of the formation of the LMC disk.
A merger event with a gas rich galaxy at about 10 - 12 Gyr could have caused this major event. The low star
formation rate of the LMC halo must have kept the chemical enrichment low enough for the RRLS to form 
during the merger. This also points to a very low stellar density halo for the LMC. If the RRLS shows the bar
feature because that they formed in the bar, then the bar was also formed during this merger event, along
with the disk. It might be possible that the globular clusters in the LMC were also formed in this merger
event. The globular clusters in the LMC show-disk like kinematics (\cite{fio83} \cite{sch92},  \cite{vb04}). Increased gas material due to the merger could have triggered the formation of the globular clusters.

We also identified that when one component of RRLS shows disk-like scale height, another component appeared to be
inflated. These two population could also indicate the early formation of the LMC. The inflated distribution
could have formed in the early part of the merger, where, the disk is not completely formed, giving it an
inflated location. If this coincides with globular cluster formation, then the RRLS in the inflated component are formed just before the
formation of the disk. The star formation must have proceeded along with the formation of the disk. 
In summary, 
we suggest that the major star formation event in the LMC happened
at $\ge$ 9 Gyr, probably with a gas rich merger. This merger resulted in the formation of most of the inner
RRLS, probably the globular clusters and the disk of the LMC. 

We also explore the possibility that all the RRLS formed in the disk and one fraction got inflated.
One scenario is where another gas-rich merger dislocated some stars
to the halo, resulting in a distribution with increased scale-height.
Another mechanism for distributing stars vertically above the plane is bar instabilities.
After the bar is formed, it thickens, its scale-height increases 
and it forms a peanut/boxy bar, due to various instabilities
(\cite{C90}, \cite{PF91}).
Once the RRLS are trapped in the bar potential, they might experience various bar 
instabilities resulting in a vertical distribution.
The fact that the observed inflated distribution in the RRLS is located
in the region of the bar could suggest a bar-related phenomenon.
If this is a viable mechanism, then the RRLS can be used to understand the
evolution of the LMC bar. 
The survival and the evolution of the bar could also be a pointer to the dark 
matter halo (\cite{mv06}).

\end{document}